\documentclass[twocolumn,prb,superscriptaddress]{revtex4-2} 

\usepackage{graphicx}
\usepackage{dcolumn}
\usepackage{bm}

\usepackage[utf8]{inputenc}
\usepackage{sidecap}
\usepackage{caption, subcaption}
\usepackage{units}
\DeclareUnicodeCharacter{2212}{-}
\usepackage[T1]{fontenc}
\usepackage{textgreek}
\usepackage{color,soul} 
\usepackage{comment}
\usepackage{float} 
\usepackage{lipsum}
\usepackage{physics}
\usepackage{array}
\usepackage{url}
\usepackage{amsmath}
\usepackage{upgreek}
\usepackage{xcolor}
\usepackage{ulem}


\begin{document}

\preprint{APS/123-QED}

\title{Low-energy excitations  in  multiple modulation-doped  CdTe/(CdMg)Te quantum wells}

\author{D.~Yavorskiy}
\affiliation{Institute of High Pressure Physics, PAS, ul. Sokołowska 29/37, 01-142 Warsaw, Poland}
\affiliation{CENTERA, CEZAMAT, Warsaw University of Technology, ul. Poleczki 19, 02-822 Warsaw, Poland}
\affiliation{Institute of Physics, PAS, al.  Lotników 32/46, 02-668 Warsaw, Poland}

\author{F. Le Mardelé}
\affiliation{LNCMI, CNRS-UGA-UPS-INSA-EMFL, 25 rue des Martyrs, 38000 Grenoble, France}

\author{I. Mohelsky}
\affiliation{LNCMI, CNRS-UGA-UPS-INSA-EMFL, 25 rue des Martyrs, 38000 Grenoble, France}

\author{M. Orlita}
\affiliation{LNCMI, CNRS-UGA-UPS-INSA-EMFL, 25 rue des Martyrs, 38000 Grenoble, France}
\affiliation{Institute of Physics, Charles University, Ke Karlovu 5, Prague, 121 16 Czech Republic}

\author{Z. Adamus}
\affiliation{Institute of Physics, PAS, al. Lotników 32/46, 02-668 Warsaw, Poland}
\affiliation{International Research Centre MagTop, Institute of Physics, PAS, al. Lotników 32/46, 02-668 Warsaw, Poland}
 
\author{T.~Wojtowicz}
\affiliation{International Research Centre MagTop, Institute of Physics, PAS, al. Lotników 32/46, 02-668 Warsaw, Poland}
\author{J. Wróbel}
\affiliation{Institute of Physics, PAS, al. Lotników 32/46, 02-668 Warsaw, Poland}
\affiliation{Institute of Applied Physics, Military University of Technology, Kaliskiego 2, 00-908 Warsaw, Poland}

\author{K. Karpierz}
\affiliation{Faculty of Physics, University of Warsaw, L. Pasteura 5, 02-093 Warsaw, Poland}

\author{J. Łusakowski}
\email[]{jerzy.lusakowski@fuw.edu.pl}
\affiliation{Faculty of Physics, University of Warsaw, L. Pasteura 5, 02-093 Warsaw, Poland}

\date{\today}

\begin{abstract}
 Low energy excitations of a two-dimensional electron gas (2DEG)  in modulation-doped multiple (ten)  quantum wells (QWs) was studied using far-infrared magneto-transmission technique at liquid helium temperatures. A large distance between neighbouring QWs of 54~nm excluded a direct interaction of electron wave functions confined  in the wells.  In four  samples which differed in the spacer width and the level of doping with iodine donors, supplied with a metallic grid coupler,  a uniform picture of exitations of the 2DEG was observed. These involved the cyclotron resonance (CR), its second harmonic (2CR) and magnetoplasmon modes (MPMs). \textcolor{black}{MPMs with a small amplitude originated from excitations of  the 2DEG in a single QW and these with a high amplitude resulted from a coherent excitation of the 2DEG in all wells. A polaron effect resulting from the interaction of the CR, 2CR and MPMs with an optical phonon was observed and discribed with appropriate models. Both types of MPMs exhibited gaps in the disperion relations at the frequency close to the 2CR, leading to Bernstein modes, which was discribed  with an appropriate (non-local) theoretical model.}  
\end{abstract}


\maketitle


\section{\label{sec:introduction}Introduction }

Low-energy  excitations of a two-dimensional electron gas (2DEG) in the far-infrared (FIR\textcolor{black}{, or THz frequencies})~\cite{Note} have been in the focus of interest in many theoretical and experimental groups since the  1950s~\cite{GDresselhaus_1955, JMLuttinger_1956, GAbstreiter_1976}.  \textcolor{black}{Here we report on FIR magneto-spectroscopy studies of modulation-doped CdTe/(CdMg)Te multi quantum wells (MQWs). In the response, we identify } several types of FIR excitations: the cyclotron resonance (CR), its second harmonics (2CR) and  magnetoplasmon modes (MPMs). In addition, we observed   an interaction of the CR, 2CR and MPMs with optical phonons (a polaron effect) and \textcolor{black}{gaps in the dispersion of MPMs in the region where it crosses the 2CR. The latter effect is a non-local phenomenon and  mimics a truly interaction between  the 2CR with MPMs. Plasma excitations in that part of MPMs spectrum are conventionally named the Bernstein modes.}   

The cyclotron resonance is probably the most characteristic excitation    of a 2DEG subjected to the magnetic field, $B$, and has been widely studied because it serves as \textcolor{black}{the main and precise tool to determie the single-particle effective mass}. \textcolor{black}{Using the quantum picture,} the basic principle of the  CR - an electric  dipole allowed transition between subsequent  Landau levels  - becomes more complex when one puts it within a frame of a general description of the response of an electron gas to arbitrary electromagnetic wave~\cite{MPGreene_1969} or one takes into account  interaction between electrons and many-body effects~\cite{ZSchlesinger_1984, CKallin_1984, CKallin_1985,AHMacDonald_1989} even though on the basic level the latter effect should not appear \textcolor{black}{in systems with a strictly parabolic band}~\cite{WKohn_1961}.

  This complexity have become  more and more important with increased quality of samples, particularly with extreme values of the electron mobility  in the case of a 2DEG. For example, in high-electron mobility GaAs-based structures one can observe nonmonotonic variations of the cyclotron effective mass with $B$ which results (at least, partially) from  filling-factor dependent screening~\cite{EBatke_1988} (see, however, a prior report~\cite{TKLee_1976}) or electron correlations observed in CR measurements which are also related to the filling factor~\cite{MManger_2001}. The CR  of composite fermions was observed as well~\cite{IVKukushkin_2002}.

  A simple one-particle approach to the CR was questioned in~\cite{YABychkov_2005, CFaugeras_2007}. It was shown that  considering  the  CR as a magnetoplasmon excitation, one can explain experimentally observed dependence of   the CR energy on $B$ and an absorption oscillator strength. Development of a time-resolved spectroscopy in the FIR domain (a technique referred to as THz-TDS~\cite{JNeu_2018}) allowed one to measure coherence times of the CR transition~\cite{XWang_2010} and establish a coherent control of this transition~\cite{TArikawa_2011}. A THz-TDS technique was also used to find out a many-body nature of decoherence of  the CR in a high-electron-mobility 2DEG in  GaAs QWs~\cite{QZhang_2014}.

Magnetoplasmon modes are natural companions of the  CR   in a solid-state plasma in bulk materials   and low-dimensional structures. They are frequently observed together with the CR, as it is also the case of the present paper. For a general reference on magnetoplasma effects in bulk materials, see Refs.~\cite{IBBernstein_1958, EDPalik_1970}. The dispersion of longitudinal waves in a 2D plasma was first considered by   Stern in 1967~\cite{FStern_1967} but it took a decade for the first observation of plasmons in a 2DEG to be reported in~\cite{SJAllen_1977}. Theoretical analysis of Stern was further developed by Greene~\cite{MPGreene_1969} and Chiu and Quinn~\cite{KWChiu_1974} to include the magnetic field. 

In the  last two decades, new directions of research of magnetoplasmons have been opened and have become dominated by studies of plasmons in graphene~\cite{FHLKoppens_2011,ICrassee_2012, ANGrigorenko_2012,GXNi_2018, LXiong_2019, MKravtsov_2024}  and  field-effect transistors~\cite{MDyakonov_1993, LVicarelli_2012,WKnap_2013, VRyzhii_2023, GRAizin_2024, JMCaridad_2024}. A particularly interesting junction of subjects: graphene, plasmons and a near-field nanoscopy can be found, e.g., in~\cite{ASoltani_2020}. Let us note that a large part of all these contemporary studies, although originating and tightly bound to fundamental physics,  are devoted to applications of plasmonic devices  as detectors or emitters of FIR radiation. 
For an older and a recent review devoted to the electron plasma in two dimensions in solids, see~\cite{MSKushwaha_2001} and ~\cite{IVZagorodnev_2023}, respectively. 

  \textcolor{black}{Cadmium telluride is a semiconductor with a strong electron - phonon interaction   which facilitates the observation of polaronic effects.}  The papers~\cite{VLGurevich_1965} and~\cite{FGBass_1966} are among the first theoretical approaches to the polaron theory. Some further contributions to the theory of the polaron effect was presented by, e.g.,  Peeters and Devreese~\cite{FMPeeters_1985} and Pfeffer and Zawadzki~\cite{PPfeffer_1988, PPfeffer_1990}. A valuable recent review  on the polaron effect is given in~\cite{CFranchini_2021} - it  contains a detailed account on the history of development of the   theory of polarons and a broad  list of references to theoretical and experimental works. An account on earlier studies can be found in~\cite{DevreeseBook}. 

 Very often,  the polaron effect is observed as an increase of the cyclotron electron effective mass when the energy of the CR transition is close to that of the optical phonon~\cite{SDasSarma_1983} as it was observed in a single CdTe/(CdMg)Te QW in a previous publication~\cite{IbGrigelionis_2015}. However, optical phonons can interact also with magnetoplasmons and  \textcolor{black}{the study} of interaction between (magneto)plasmons and optical phonons is another broad area of research in solids. Theoretically, in the case of a degenerate statistics, this interaction was first described by Yokota~\cite{IYokota_1961}, Varga~\cite{BBVarga_1965} and  Lee and Tzoar~\cite{YCLee_1965}  and then observed by Mooradian and Wright~\cite{AMooradian_1966}. Dispersion of plasmon-phonon polaritons   can be determined by Raman scattering, which is the most widely used technique in such studies,  or by photoluminescence which is particularly useful  in the case of magnetoplasmons (see, e.g.,~\cite{PPerlin_1995} and~\cite{AWysmolek_2006}, respectively, reporting results on bulk semiconductors). The interaction of plasmons with optical phonons in two-dimensional systems is considered theoretically in many papers  (e.g., for a graphene-oriented review, see~\cite{SSXiao_2016})  but not many experimental results have been so far presented for a 2DEG in semiconductor QWs~\cite{LVKulik_2000}.

 The present paper shows a strongly polar   2D semiconductor  system in which such an interaction can be observed. \textcolor{black}{A conclusion  which results from our phenomenological analysis of the polaron effect in samples studied is that a quantitative description of this effect, which is in our case defined by  the frequency $\omega_{ph}$ of a phonon taking part in the polaron interaction, seems to  depend on the sample.  Tentatively, we  attribute this effect to sample-dependent  disorder of the layers adjacent to the QWs.}

  \textcolor{black}{Derivation of the dispersion  relation of the plasma oscillations requires an approach in which one first determines the conductivity tensor of the plasma and then uses it to solve the Maxwell's equations.  Generally, this leads to the plasma  dispersion relation which is non-local, i.e., next to a an explicit dependence on the plasmon wave vector it also depends on the product of the    wave vector $k$ and the cyclotron radius $r_c$}. Folowing this procedure, Bernstein showed~\cite{IBBernstein_1958} that  at given magnetic field, the gaps open at frequencies near the harmonics of the CR.  \textcolor{black}{Thus}, when the frequency of a MPM approaches that of a CR harmonic, an avoided crossing  appears in the MP dispersion. This phenomenon attracted attention of researchers and was studied in a large number of papers, both theoretical and experimental, an exemplary account of which are given by~\cite{CKNPatel_1968, NTzoar_1969, FABlum_1970, NJHoring_1972, TKamimura_1978,RRSharma_1980,EBatke_1985, VGudmundsson_1995,DEBangert_1996,JLefebvre_1998,RKrahne_2000,SHolland_2003,AWysmolek_2006, AAKapustin_2021}.

 \textcolor{black}{Traditionally, the parts of the non-local plasmon dispersion   which are situated in the vicinity of the  harmonics of the CR are called the Bernstein modes.    As it is clearly shown, e.g, in Ref.~\cite{KWChiu_1974, AAKapustin_2021}, the upper hybrid mode is composed of these  parts of the  dispersion relation for which non-local corrections are negligible and the dispersion can be very well approximated by the local approach. It follows that a division of the MP dispersion to the upper hybrid mode and Bernstein modes is artificial and particularly misleading when these two notions are confronted as entities of a different origin.  However, for brevity and following the tradition, we will use in the following the notion "Bernstein modes" by which we will mean just a part of the dispersion    in the vicinity of the harmonics of the CR.  We will come back to an analysis of the non-local dispersion relation in Section~\ref{sec:discussion}. As a perfect introduction to the local  theory of plasma waves, we recommend in addition to \cite{EDPalik_1970} also introductory chapters of \cite{StixBook}.}

Interestingly, the Bernstein modes have so far been studied almost exclusively in bulk GaAs and GaAs-based low dimensional structures. One of the first observations of Bernstein modes in other solid-state systems, GaN/(AlGa)N heterostructures,  was reported in~\cite{KNogajewski_2013} which was followed by a recent observation in graphene~\cite{DABandurin_2022}. The present paper  \textcolor{black}{presents an analysis of  Bernstein modes in  CdTe/(CdMg)Te MQWs. Here, according to our interpretation, the gaps in the dispersion relation open in the case of plasmons excited in a single QW and  coherently in all ten QWs. The  former type of Bernstein modes has not yet been observed, to the best of our knowledge. }   

The present paper \textcolor{black}{considers results of}  a FIR Fourier transmission spectroscopy of CdTe/(CdMg)Te MQWs. The case of  single QWs was considered in~\cite{IaGrigelionis_2015, IbGrigelionis_2015} where one can find a referenced summary on previous  studies on the CR in CdTe/(CdMg)Te  single QWs. 

The paper is organized as the following. Section \ref{sec:experimental} describes the samples used and the experimental system. Results of measurements are presented in Section~\ref{sec:results} and discussed in  Section~\ref{sec:discussion}. Then we conclude the paper.

\section{\label{sec:experimental}Experimental}

The samples studied were grown by the molecular beam epitaxy on a semi-insulating GaAs substrate and  contained ten QWs. In each case, the QWs   were grown from CdTe  while barriers were always a mixed crystal Cd$_{0.7}$Mg$_{0.3}$Te. All  QWs in a given sample were nominally identical - their width, the thickeness of the spacer $d$ and the doping was the same (doping was introduced after a QW was grown). The level of doping is described by the number of monolayers ($N$) in which iodine donors were introduced. The temperature of the iodine source was the same during the growth of all doped layers in all samples which allows us to assume that the concentration of donors in a doped region is the same.  The samples differed by $N$ and $d$. The distance between subsequent QWs is equal to   54~nm which allows us to consider them as non-interacting in the sense that electron's wave functions from neighbouring wells  do not overlap. However, interaction with  FIR radiation which leads to a collective excitation of all QWs in the sample is evidenced by experimental data, as it will be shown further on.

\begin{figure}
\includegraphics[scale=3.0]{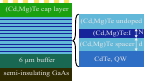}
\caption{\textcolor{black}{A scheme of samples' structure (not to scale). One period of the active part of the sample is enlarged to show the sequence of the layers and indicate the meaning of parameters  $d$ and $N$. } }
\label{SamplesScheme}
\end{figure}

 \textcolor{black}
{A scheme of samples' structure is shown in Fig.~\ref{SamplesScheme}.} Parameters of the samples are given in Table~\ref{tab:alfa}; the last column shows the value of the electron effective mass (in  units of the free electron mass, $m_e$) determined from the CR by a Fourier spectroscopy  to be described further on; the relative error of $m^*$ is equal to about 10$^{-3}$. The values of concentration $n$ (per well) presented in Table~\ref{tab:alfa} are obtained from fitting of magnetoplasmon dispersions, which is explained in Section~\ref{sec:discussion}.  As one can notice, these values of $n$  change with technological parameters $d$ and $N$   as expected (i.e., the thinner spacer and the wider doped layer, the higher the electron concentration). 

Two sets of samples were studied: one set contained   as grown samples and the other -- samples with a litographically prepared metallic (Cr/Au) 50~nm-thick grid on their surface. In all cases,  the period of the grid was 2~$\mu$m with  $\alpha = 0.5$ of the geometrical aspect ratio.

\begin{table}[ht!]
        \caption{Samples' parameters }
	\begin{center}
	\begin{tabular}{|c|c|c|c|c|c|}
		\hline
		Sample & $d$  [ nm ] & $N$ & n  [ cm$^{-2}$ ]   & m$^*$ [m$_e$]\\
		\hline

            MQW$_1$ & 5 & 12 & 1.10$\times$10$^{12}$ &  0.103 \\
            \hline
            MQW$_2$ & 10 & 12 & 1.07$\times$10$^{12}$ &  0.103 \\
            \hline
            MQW$_3$ & 20 & 12 & 9.61$\times$10$^{11}$ & 0.101 \\
            \hline
  	MQW$_4$ & 20 & 9 & 8.69$\times$10$^{11}$ &  0.100 \\
            \hline
	\end{tabular}
	\label{tab:alfa}
	\end{center}
\end{table}

The samples were placed in the center of a 16~T coil and cooled in   dark  to 4.2~K with helium exchange gas at low pressure. No additional near-band-gap illumination during measurements was used.  Transmission spectra were registered with a Fourier spectrometer  as a function of $B$ with a bolometer placed below the sample. The spectra were registered at selected values of $B$ each 0.25~T with the resolution of about  1.0~cm$^{-1}$. 

\begin{figure}
\includegraphics[scale=0.55]{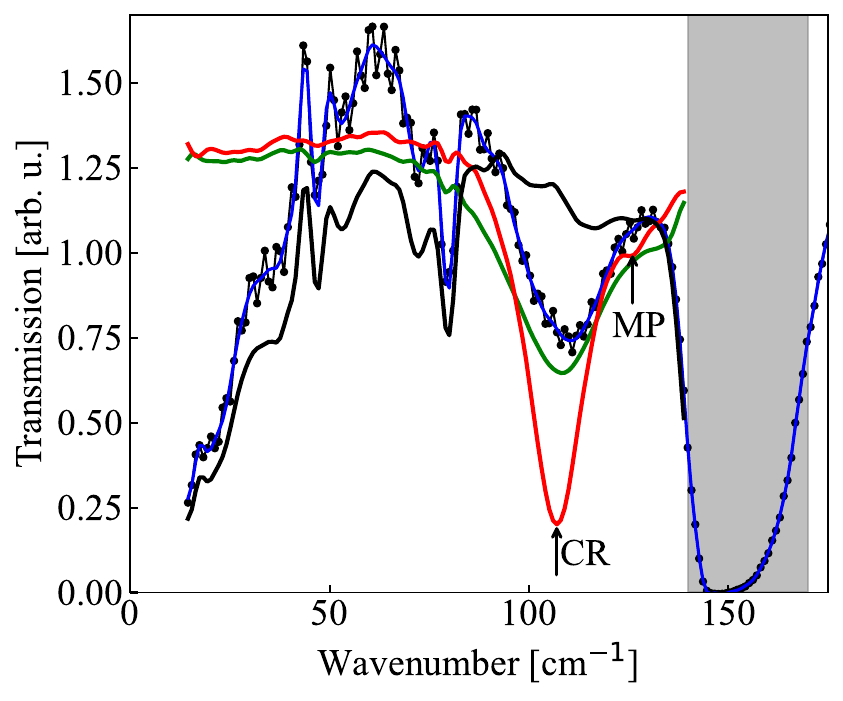}
\caption{ \textcolor{black}{Black dots: the  raw spectrum measured on MQW$_1$ at 12 T. Blue: the same spectrum after removing Fabry-Perot interferences. Solid black: a mean spectrum. Solid green: the blue spectrum after normalization with the mean spectrum. Solid red: the result of treatment of a spectrum measured at 12 T on GMQW$_1$. The arrows show positions of the cyclotron resonance (CR  - the same for the green and red spectra) and the magnetoplasmon  (MP, present in GMQW$_1$ only). The grey rectangle shows the Reststrahlen band of CdTe, between 140 and 170 cm$^{-1}$.}}
\label{Example}
\end{figure}

\textcolor{black}{A graphical presentation of the procedure of numerical data treatment is presented in Fig.~\ref{Example} for spectra measured at 12~T on MQW$_1$ and GMQW$_1$ (the letter G stands for the grid).}
The raw data  \textcolor{black}{(black dots in Fig.~\ref{Example})} were first numerically filtered to remove Fabry-Perot interferences. \textcolor{black}{ As a normalization procedure, we chose to divide each filtered spectrum by a mean spectrum which was obtained as the arythmetic average of all filtered  spectra measured for given sample in the whole range of $B$   (solid black line in    Fig.~\ref{Example}).  The results are shown as green and red solid lines. This allowed us to determine positions of the main spectral features, i.e., the CR and MPs, as is indicated by arrows in Fig.~\ref{Example}. However, to analyze spectral features that are weaker in their intensity, it was necessary to construct color maps showing}  a derivative of the signal with respect to the magnetic field (by the derivative we mean a difference of two  spectra measured at $B$ differed by 0.25~T; \textcolor{black}{in calculating the derivative, we used spectra after filtering-out Fabry-Perot interferences but not normalized}). The high-energy cut-off of the data presented   at about 140~cm$^{-1}$  \textcolor{black}{is defined by}  the Reststrahlen band of CdTe.

\section{\label{sec:results}Results  }

\begin{figure*}[ht!]
     \centering
         \includegraphics[width=1.\textwidth]{ 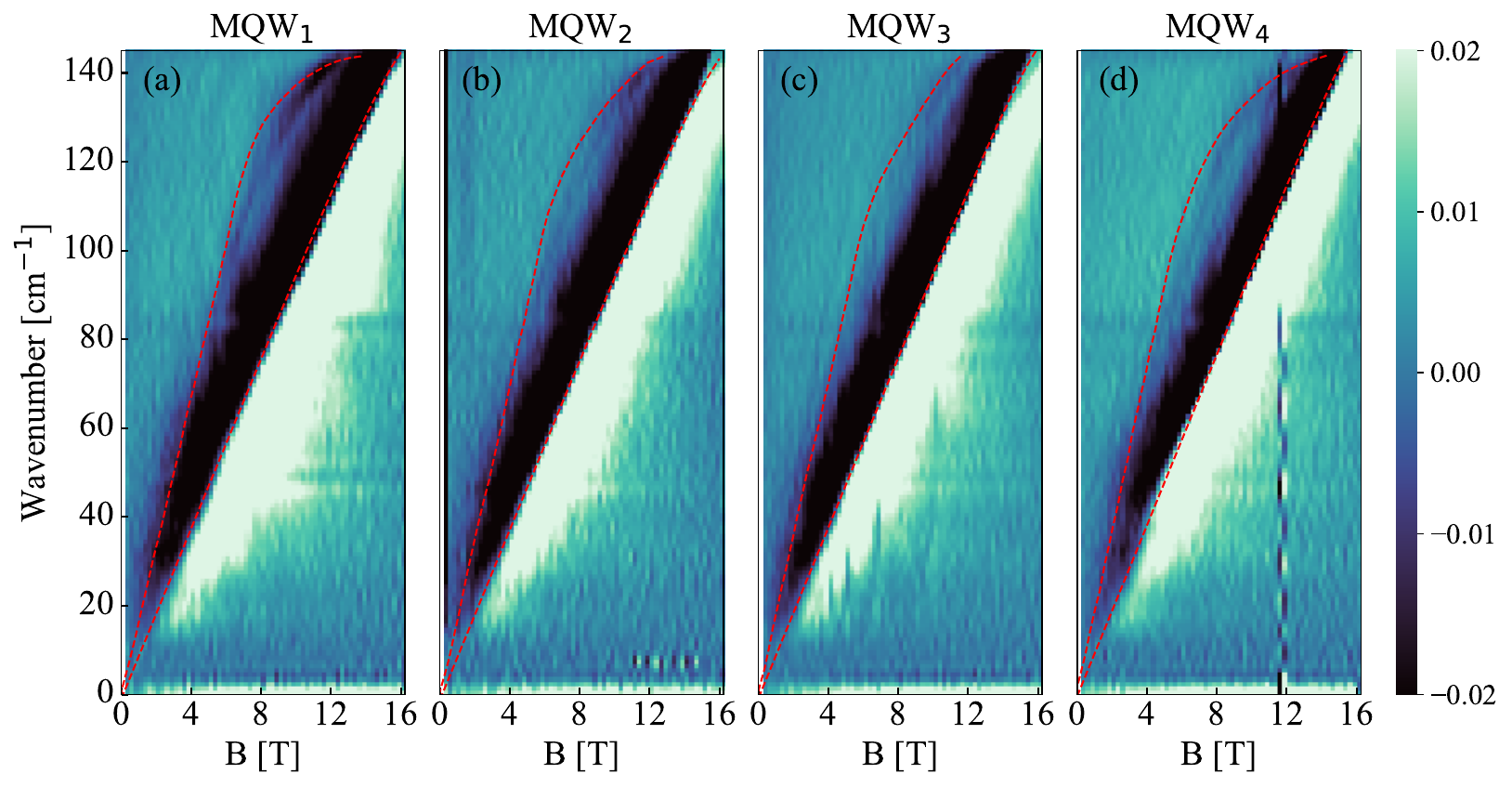}
	\caption{Transmission data  on MQW$_1$- MQW$_4$ samples (a) - (d), respectively. Thin red dotted lines show the CR and the 2CR.  }
	\label{fig:ALLCR}
\end{figure*}

 Figure~\ref{fig:ALLCR} shows results measured on MQW samples without a grid. The dotted red lines show  the  expected ($\hbar eB/m^*$)  first and the second  harmonics of the CR.  There are two clearly visible features: a small deviation from the linearity caused by the polaron effect in the case of the CR and a  much stronger polaron interaction in the case of the 2CR. 

\begin{figure*}[ht!]
     \centering
         \includegraphics[width=1.\textwidth]{ 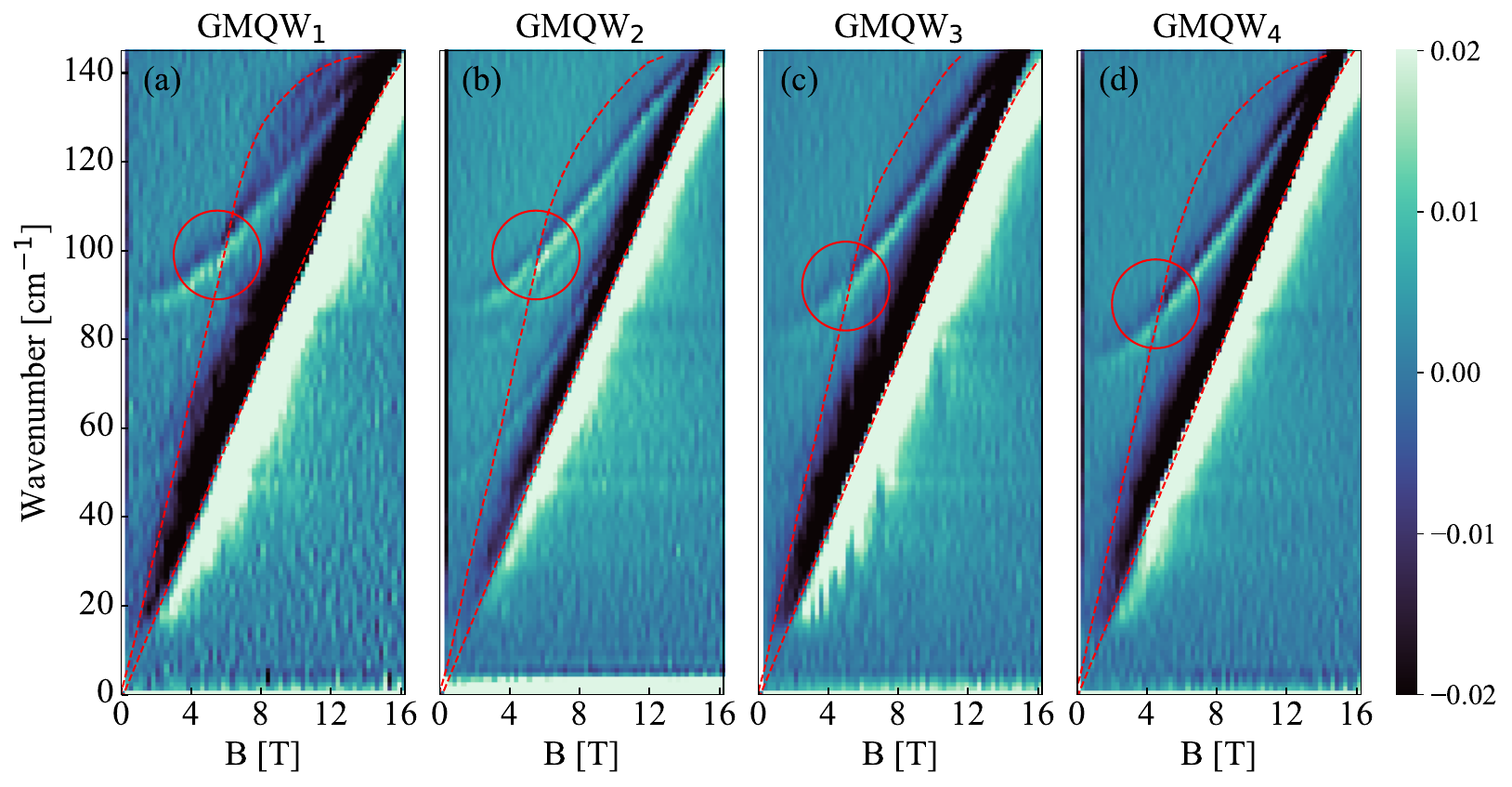}
	\caption{Transmission data  on GMQW$_1$ - GMQW$_4$ samples (a) - (d), respectively. The red circle shows the region of Bernstein modes.}
	\label{fig:ALLMP}
\end{figure*}

Transmission results on samples with a grid are presented in Fig.~\ref{fig:ALLMP}. \textcolor{black}{As compared to Fig.~\ref{fig:ALLCR}, the samples with a grid display considerably richer behaviour.}  First,  the interaction of the 2CR with an optical phonon is less pronounced,  however, an additional feature emerges -    \textcolor{black}{gaps in the magnetoplasmon dispersion} which are clearly visible in   areas indicated by  the red circles. Second, there is a strong MPM   with the energy  about 80~cm$^{-1}$ at $B=0$ which also interacts with the optical phonon. \textcolor{black}{This interaction leads to a bending of the MPM   towards the CR line.} Lower-energy MPMs are well visible in the case of GMQW$_2$ sample, although they could be also detected in the data of other samples under a careful inspection. 

\begin{figure*}[ht!]
     \centering
         \includegraphics[width=1.0\textwidth]{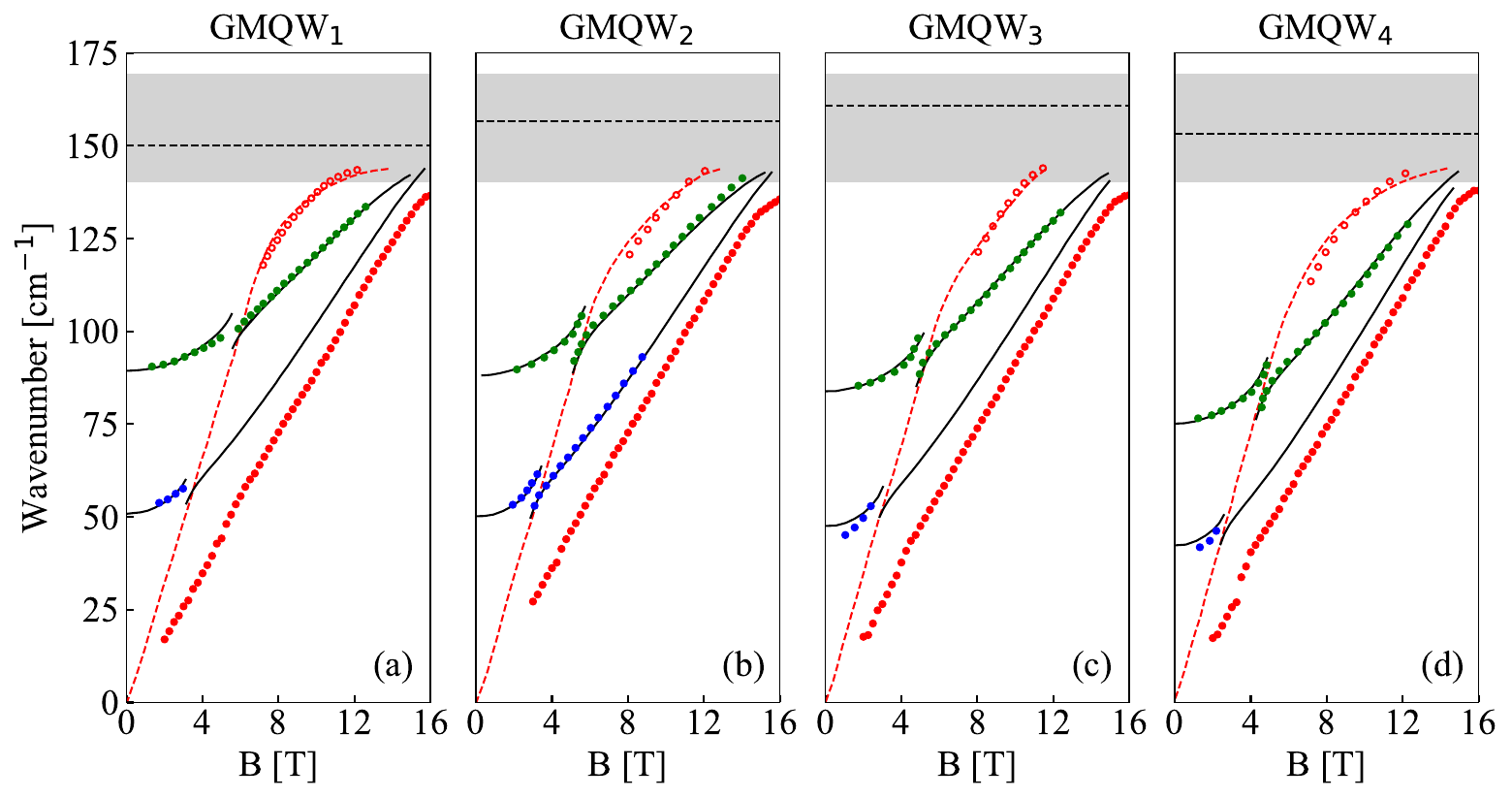}
	\caption{Red solid dots: CR with interaction with an optical phonon.  Red open dots: 2CR with interaction with an optical phonon.  Red dashed lines: results of fitting of Eq.~\ref{eq:polaron}.  Blue dots: a low-energy magnetoplasmon with interaction with 2CR. Green dots: a high-energy magnetoplasmon with interaction with 2CR.   Black   lines: calculated magnetoplasmon energy according to Eq.~\ref{eq:nlmagpl}. Grey rectangle - the  CdTe Reststrahlen band. Dashed lines - fitted energy of phonons $\omega_{ph}$ from  Table~\ref{tab:polaron}.  }
	\label{fig:MODEL}
\end{figure*}

To compare experimental results with a theoretical model, we present in Fig.~\ref{fig:MODEL} the positions of the above mentioned spectral features.

\section{\label{sec:discussion}Discussion}

\subsection{The second harmonics of   the CR and polaron interactions}

 For a uniform system, electric-dipole allowed optical transitions between Landau levels (LLs) are possible only between  pairs of adjacent LLs. The second or higher harmonics of the CR were  observed in some optical FIR or microwave experiments~\cite{MAZudov_2001, YDAi_2010,  MBialek_2015, MLSavchenko_2021}  but no consistent picture of conditions at which the 2CR may appear has so far been established.  Generally, one can expect that symmetry-breaking mechanisms must be involved like a strong scattering, electric field or non-parabolicity.

Particularly appealing theoretical models were proposed in~\cite{SAMikhailov_2011,VAVolkov_2014} which relate excitation of the CR harmonics to a strong electric field generated by the incoming electromagnetic wave on edges of metallic elements (ohmic contacts, gates, grids)  present on the sample's surface.  However, our data seem to show just the opposite: the 2CR in samples with the grid is much weaker than in samples without it. This means that other symmetry-breaking mechanism should be invoked. 
 
  In the case of samples studied \textcolor{black}{here}, the 2CR itself is not only clearly visible in the data but   also  exhibits  evidence of interaction    with MPMs  (blue and green dots in Fig.~\ref{fig:MODEL})  and the  optical phonon (red   points in Fig.~\ref{fig:MODEL}). 

   CdTe is a material with a strong interaction of electrons with optical phonons.  This is expressed by a high value of the Fr\"{o}hlich constant equal to 0.286~\cite{RJNicholas_1992}. The polaron effect was observed in single CdTe/(CdMg)Te QWs in the past~\cite{YImanaka_1998, GKarczewski_2002,IaGrigelionis_2015} as a deviation from a linear dependence of the   CR transition at high $B$. Here,  the polaron interaction \textcolor{black}{is manifested via $B$-dependence of }  the CR,  2CR as well as on MPMs. 

 To describe the interaction of the CR and 2CR with optical phonons, we apply a phenomenological approach based on a Hamiltonian $\hat{H}$ of a two-level system with an interaction described by $\Omega_i$:
\begin{equation} 
\hat{H} = \hbar\left[\begin{array}{cc}
\omega_{i} & \Omega_{i}  \\
\Omega_{i}^*  & \omega_{ph} 
\end{array}
\right].
\label{eq:polaron}
\end{equation}
The above Hamiltonian describes interaction of two oscillators:  a CR-related oscillator (with $i$ =   1 or 2 corresponding to the  CR or   2CR, respectively)  with the frequency $\omega_{i} = i eB/m^*$ and an optical phonon with the frequency $\omega_{ph}$.
The energy of interacting levels is then given by
\begin{equation}
E_i =  \frac{\hbar}{2}\left(\omega_{i} + \omega_{ph} \pm \sqrt{(\omega_{i} - \omega_{ph})^2 + 4 |\Omega_i|^2} \right)
\end{equation}
We fitted  the lower energy solution to experimental values of the  2CR frequency dependence on $B$ in the case of samples without and with the grid obtaining mean values of $|\Omega|_{2CR}$ and $\omega_{ph}$.  Next, the extracted parameters $\omega_{ph}$ were used to obtain  the  values of $|\Omega|_{CR}$  describing the polaron interaction in the case of  the first harmonic.    Results of the fitting are presented in Table~\ref{tab:polaron}  while the dependencies resulting from the fitting are shown as red dashed lines in Fig.~\ref{fig:MODEL}. \textcolor{black}{This figure shows also the Reststrahlen band of CdTe and the energy of phonons determined by fitting.  The Ockham's razor allows us to assume that both the CR and 2CR are influenced by the same phonon, but this does not necessarily need to be the case.  }

The presence of interfaces modifies energies of phonons  known \textcolor{black}{from studies of} bulk materials. First, \textcolor{black}{there are} phonons propagating in the proximity of interfaces, which are absent in bulk materials: these vibrations involve different  sets of atoms than in bulk which changes their frequency. Such   interface-related phonons are particularly important in the case when the polaron interaction involves a 2DEG residing in a proximity of the interface.  Second factor, especially important in a strongly  polar material such as CdTe,  is due to modification of electron-phonon interaction   resulting  from localization of electron wave function in QWs which changes  efficiency of screening. For a general reference on this subject we recommend Ref.~\cite{Ridley} and Ref.~\cite{DLiu_2018,AMdePaula_1998} specifically devoted to studies of phonons in CdTe-based nanostructures.  Photoluminescence spectra with phonon replicas and Raman scattering spectra obtained  on MQW$_2$ sample were presented in~\cite{WSolarska_2023}. It was shown that the frequency of optical phonons which result from such processes were shifted with respect to values obtained in studies on bulk CdTe. 
For these reasons, we   do not specify the nature of the phonon which interacts with the CR and 2CR. A final identification should be based on   more complete Raman scattering - oriented studies on these samples which is beyond the scope of the present paper. 

Landolt-B\"{o}rnstein data base~\cite{Strauch_2012} reports  values of a transverse optical phonon in CdTe between 141 and 146~cm$^{-1}$. Analyzing the values of $\omega_{ph}$ we notice that at the constant $N=12$ there is a systematic shift to higher frequencies with increasing $d$ but then a rapid decrease for the smallest doping (the largest $d$ and the lowest $N$, sample $l=$4). 
  Also, we note that that the polaron effect on the 2CR  is essentially stronger than in the case of the CR. This can be observed directly in experimental data where bending of the 2CR line occurs at  much lower $B$ than in the case of the CR. This fact is reflected in the value of the interaction constant $|\Omega|_{2CR}$ which is clearly larger than $|\Omega|_{CR}$.

 \textcolor{black}{At the moment, we cannot give a definitive interpretation of this difference. However, if the strenght of the interaction is to be estimated  on the basis of a deviation of the CR (or 2CR) vs $B$ dependence from the straight line, then we may refer to Fig. 2b in Ref.~\cite{IbGrigelionis_2015} which shows the polaron effect on the CR in a single CdTe/(CdMg)Te QW with parameters very similar to these of QWs studied here. The polaron effect there observed starts to be visible  at about 10~T and is essentially stronger than that observed in the present study, where it is visible only at the highest $B$ about 14~T. The conclusion is that the polaron effect in the CR  in   CdTe/(CdMg)Te QWs is apparently very sensitive to microscopic details of the QWs studied. This could be related to the disorder of the layers or interfaces adjacent to the QW modyfying the spectrum of phonons interacting with electrons in the QWs. Different phonon frequencies $\omega_{ph}$ obtained from fitting  suggest that these values are sample-specific which could support this disorder - related interpetation. }

 \begin{table*}[ht!]
       \caption{Polaron interaction - average values from MQW$_l$ and GMQW$_l$ samples; $ l = 1, \ldots, 4$}
	\begin{center}
 
	\begin{tabular}{|c|c|c|c|c|c|}
		\hline
		$l$ & $d$  [ nm ] & Doping  [ ML ] & $ \hbar|\Omega|_{CR}$  [ cm$^{-1}$ ] & $\hbar|\Omega|_{2CR}$   [ cm$^{-1}$ ]   &$ \omega_{ph}$ [cm$^{-1}$  ]\\
		\hline         
      1 & 5 & 12& $1.68 \pm 0.03$    & $3.15 \pm 0.07$ & $150.1 \pm 0.6$ \\
            \hline
          2 & 10 & 12& $ 1.70 \pm 0.03 $  & $4.22\pm 0.12$ &  $156.7 \pm 1.3$ \\
            \hline
           3 & 20 & 12 & $ 1.58 \pm 0.05 $   & $4.22 \pm 0.05$ & $160.7 \pm 0.6$ \\
            \hline
  	4 & 20 & 9 & $1.26  \pm 0.06  $  & $3.79 \pm 0. 08$  & $153.3 \pm 0.8$ \\
            \hline
	\end{tabular}
\label{tab:polaron}

\end{center}
\end{table*}

\subsection{Magnetoplasmon  modes}
  Magnetoplasmon modes are clearly visible in data presented in Fig.~\ref{fig:ALLMP} and their dispersion exhibits two features. The first one is an avoided crossing at the energy where the magnetoplasmon dispersion coincides  with   the 2CR transition, \textcolor{black}{which is traditionally referred to  as Bernstein modes, as it was explained in Introduction.} The second one is  bending of the magnetoplasmon dispersion resulting from interaction of magnetoplasmons with the optical phonon.    Both interactions can be modeled with the dispersion relation given by: 
\begin{equation}
\frac{\omega^{2}-\omega_{LO}^{2}}{\omega^{2}-\omega_{TO}^{2}}-\frac{\omega_{p,j}^{2}}{X_{j}^{2}}\sum_{l=1}^{\infty}\frac{4l^{2}J_{l}^{2}(X_{j})}{\omega^{2}-(l\omega_{c})^{2}}=0,
\label{eq:nlmagpl}
\end{equation}
where $\omega_{p,j}$ and $k_j$  is the frequency and the wave vector of $j^{\mbox{\tiny{th}}}$ plasmon mode at $B=0$, respectively:
\begin{equation}
\omega_{p,j} = \sqrt{\frac{e^2n k_j}{2m^*\epsilon_0\epsilon_{\mathrm{eff}}}},
\end{equation} 
and $X_j = k_j r_c$  where $r_c$ is the cyclotron radius. For derivation of Eq.~\ref{eq:nlmagpl} see~\cite{IaGrigelionis_2015} and references therein; for short - this formula involves polarization of the system coming from the lattice   (the first term) and from a non-local effect in  the electron gas (the second term). 

\textcolor{black}{In the analysis of the dispersion relation we are interested in the long-wavelengh limit with $X_j \ll1$. Results of numerical calculations of the disperison relation resulting from the second term in Eq.~\ref{eq:nlmagpl} are shown in \cite{AAKapustin_2021} while a  discussion of analitical expressions is presented in \cite{KWChiu_1974}, Sec. IVA. Following the latter reference, one can show that if $\omega/\omega_c$ is not too close to 2 (and also to any integer greater than 2), then the dispersion relation resulting from Eq.~\ref{eq:nlmagpl}  coincides with that based on local classical considerations, i.e., with the upper hybrid mode.
A further   analysis of the dispersion at $\omega/\omega_c$ close to any integer equal or greater than 2  shows that the dispersion splits and the gaps in the dispersion open what  is shown by numerical calculations   in \cite{AAKapustin_2021}. 
}

The effective dielectric function $\epsilon_{\mathrm{eff}}$ is different for gated and ungated plasmons. In the case of a grid-gated samples, one deals with a partially gated surface. In  previous publications, where magnetoplasmons were studied in grid-gated samples  of  CdTe/(CdMg)Te QWs  (Ref.~\onlinecite{IaGrigelionis_2015}), GaAs/(GaAl)As  (Ref.~\onlinecite{MBialek_2015})  and GaN/(GaAl)N (Ref.~\onlinecite{KNogajewski_2013}) heterostructures, we found that  the frequency of magnetoplasmon resonances  were very well reproduced if one took a weighted average of dielectric functions of gated ($\epsilon_g$)  and ungated ($\epsilon_{ug}$) plasmons and used a geometrical factor ($\alpha$) describing the percentage of surface covered with the metal to obtain $\epsilon_{\mathrm{eff}} = \alpha \epsilon_g + (1-\alpha) \epsilon_{ug}$. In this formula, $\epsilon_g = 1/2[\epsilon_s + \epsilon_b\coth(k_jd)] $  and $\epsilon_{ug} = 1/2[\epsilon_s+ \epsilon_b(1+\epsilon_b \tanh(k_jd))/(\epsilon_b + \tanh(k_jd))]$~\cite{DVFateev_2010}, where $\epsilon_s$ and $\epsilon_b$ are dielectric constants of the QW and the barrier, respectively.

Solutions of   Eq.~\ref{eq:nlmagpl} are shown as   black lines in Fig.~\ref{fig:MODEL}. The fitting procedure leading to these dependencies started with the electron effective mass  determined from the CR shown in Fig.~\ref{fig:ALLCR} taking the data between 6 and 12~T only  to avoid the noise in the signal at low-$B$ and the polaron interaction at the  highest $B$. The values of the dielectric constant of CdTe (equal to 7.1) and that of (CdMg)Te (equal to 5.9) as well as optical phonons ($\omega_{TO}$ = 140.1~ cm$^{-1}$ and $\omega_{LO}$ = 169.45~ cm$^{-1}$) were the same as used in~\cite{IaGrigelionis_2015}.  

As  evident from the above description, the dispersion of magnetoplasmons involves  many parameters and one has to decide which one is the best  \textcolor{black}{to reproduce the experimental data theoretically.}  There are two natural candidates: the effective dielectric constant and the electron concentration. As we explained above, we verified on different semiconducting systems the validity of using the effective dielectric constant in the form of a weighted average of constants of gated and ungated plasmons. Therefore, in this study we  follow this approach and we choose  the electron concentration as  a fitting parameter.

There are two MPMs  clearly visible for each data set presented in Fig.~\ref{fig:MODEL} (blue and green symbols) with the higher-energy mode   stronger as compared  to the lower-energy one.  As a starting point for the fitting, we used results of our magneto-transport experiments (not presented in this paper) which allowed us to estimate a 2DEG concentration in each sample. Then we  plotted $\omega_{p,j}$ for plasmonic modes, numbered by $j$, propagating in a single QW. It appeared that  in each sample the low-frequency mode   nearly coincided with $j$=2 plasmonic mode in a single QW. At this point, the electron concentration was finally adjusted to reproduce the data with the dispersion of $j$=2 mode. The resulting concentrations are presented in Table~\ref{tab:alfa}. As we mentioned above, a close inspection of the data allows one to find weak signatures of the fundamental $j$=1 mode.  Having identified the low-frequency mode, we found that the other one corresponds to a fundamental mode of a system which can be described as a single QW but with the electron concentration ten times higher than in a corresponding single QW. Thus, we conclude that there are arguments showing that the  high-frequency mode results from a coupled response of all QWs in the sample. A high intensity of this mode supports this interpretation. 

On the other hand, we would like to point out that the above conclusion is an experimental one and requires a theoretical support coupled to yet another type of experiment, e.g., Raman scattering. Electrodynamics of a layered electron gas was considered first by Fetter~\cite{ALFetter_1973, ALFetter_1974}  then by other authors (see, for example~\cite{SDasSarma_1982, GFGuliani_1983,   JKJain_1985})  and numerical simulations were presented by Aleshkin and Dubinov~\cite{VYAleshkin_2022}. Experimentally, the plasma dispersion was studied in such systems by inelastic light scattering~\cite{DOlego_1982, GFasol_1986}. The theory predicts that the layered 2DEG behaves as a 2D system for large separation of the QWs, i.e., when $k d \gg 1 $, otherwise its dispersion tends to that of a bulk material. In our case, the wave vector, defined by the period of the grid is $k = 2 \pi /2$~$\mu$m$^{-1}$, $d \approx$ 50~nm and $kd \approx 0.15$ which is on the border between these two cases.  Thus we put off the final conclusions on the nature of the strong magnetoplasmon mode until Raman scattering epxriments will have been performed.

The polaron interaction of MPMs is less pronounced as compared to the case of the CR and 2CR although experimental points follow   well the dependence given by Eq.~\ref{eq:nlmagpl} (black   lines in Fig.~\ref{fig:MODEL}). The reason for the absence of a clear bending of this dispersion comes from the fact that magnetoplasmons interact with longitudinal optical phonons which frequency is about 170~cm$^{-1}$~\cite{Strauch_2012} and is far above the range of energy where magnetoplasmon modes were presented  in Fig.~\ref{fig:ALLMP}.

\section{\label{sec:conclusions}Conclusions}
 In conclusion, multiple CdTe/(CdMg)Te modulation-doped QWs proved to be a semiconducting system which allows one to observe a series of low-energy excitations of a 2DEG in a single magneto-spectroscopy experiment. These include: the CR with its second harmonic, magnetoplasmons and interaction of these excitations with optical phonons showing the polaron effect. Also, the  gaps in the dispersion of MPMs  leading to the  Bernstein modes, were observed in CdTe-based  structure for the first time.  A particularly interesting observation was a high-amplitude MPM  which was interpreted as resulting from a coherent excitations of oscillations of the electron plasma  in all ten QWs. \textcolor{black}{The Bernstein modes were observed also for this coherent mode (for the first time, to the best of our knowlege) which underlines  the validity of a  general non-local description of plasma excitations. }

\begin{acknowledgements}

This research was partially supported by the Polish National Centre grant UMO-2019/33/B/ST7/02858 by the “MagTop” project (FENG.02.01-IP.05-0028/23) carried out within the “International Research Agendas” programme of the Foundation for Polish Science co-financed by the European Union under the European Funds for Smart Economy 2021-2027 (FENG). Publication subsidized from the state budget within the framework of the programme of the Minister of Science (Polska) called Polish Metrology II project no. PM-II/SP/0012/2024/02, amount of subsidy 944,900.00 PLN, total value of the project 944,900.00 PLN.

The work was supported by the European Union through ERC-ADVANCED grant TERAPLASM (No. 101053716). Views and opinions expressed are, however, those of the author(s) only and do not necessarily reflect those of the European Union or the European Research Council Executive Agency. Neither the European Union nor the granting authority can be held responsible for them. We also acknowledge the support of "Center for Terahertz Research and Applications (CENTERA2)" project (FENG.02.01-IP.05-T004/23) carried out within the "International Research Agendas" program of the Foundation for Polish Science co-financed by the European Union under European Funds for a Smart Economy Programme.

The analysis of magnetotransport data by E. Imos is also acknowledged. 

\end{acknowledgements}




\bibstyle{plain}
\bibliography{lit}

\end{document}